\documentclass[]{spie}  

\def\msun{\rm M_{\odot}}

\def\lsun{\rm L_{\odot}}

\def\kms    {\ifmmode{{\rm ~km~s}^{-1}}\else{~km~s$^{-1}$}\fi}
\def\lsun   {\ifmmode{{\rm ~L}_\odot}\else{~L$_\odot$}\fi}
\def\msun   {\ifmmode{{\rm ~M}_\odot}\else{~M$_\odot$}\fi}

\newcommand{\simle}{\mbox{$\stackrel{<}{_{\sim}}$}}

\def\arcdeg{\hbox{$^\circ$}}

\usepackage{wrapfig}
\usepackage{amsmath,amsfonts,amssymb}
\usepackage{graphicx}
\usepackage[colorlinks=true, allcolors=blue]{hyperref}

\title{Prospects for using drones to test formation-flying CubeSat concepts, and other astronomical applications}

\author[a]{John D. Monnier}
\author[a]{Prachet Jain}
\author[a]{Mayra Gutierrez}
\author[a]{Chi Han}
\author[a]{Sara Hezi}
\author[a]{Shashank Kalluri}
\author[a]{Hirsh Kabaria}
\author[a]{Brennan Kompas}
\author[a]{Vaishnavi Harikumar}
\author[a]{Julian Skifstad}
\author[a]{Janani Peri}
\author[a]{Emmanuel Hernandez}
\author[a]{Ramya Bhaskarapanthula}
\author[a]{James Cutler}
\affil[a]{University of Michigan, Ann Arbor, MI, USA}

\authorinfo{Further author information: (Send correspondence to J.D.M.)\\J.D.M.: E-mail: monnier@umich.edu}

\begin{document} 
\maketitle

\begin{abstract}
Drones provide a versatile platform for remote sensing and atmospheric studies.  However, strict payload mass limits and intense vibrations have proven obstacles to adoption for astronomy. We present a concept for system-level testing of a long-baseline CubeSat space interferometer using drones, taking advantage of their cm-level xyz station-keeping, 6-dof freedom of movement, large operational environment, access to guide stars for end-to-end testing of optical train and control algorithms, and comparable mass and power requirements.   
We have purchased two different drone platforms (Aurelia X6 Pro, Freefly Alta X)
and present characterization studies of vibrations, flight stability, gps positioning precision, and more. We also describe our progress in sub-system development, including inter-drone laser metrology, realtime gimbal control, and LED beacon tracking.
Lastly, we explore whether custom-built drone-borne telescopes could be used for interferometry of bright objects over km-level baselines using vibration-isolation platforms and a small fast delay for fringe-tracking.
\end{abstract}


\section{INTRODUCTION}
\label{sec:intro}  
\begin{wrapfigure}[19]{r}{3.6in}
\centering
\vspace{-0.05in}
\mbox{
\includegraphics[width=3.6in]{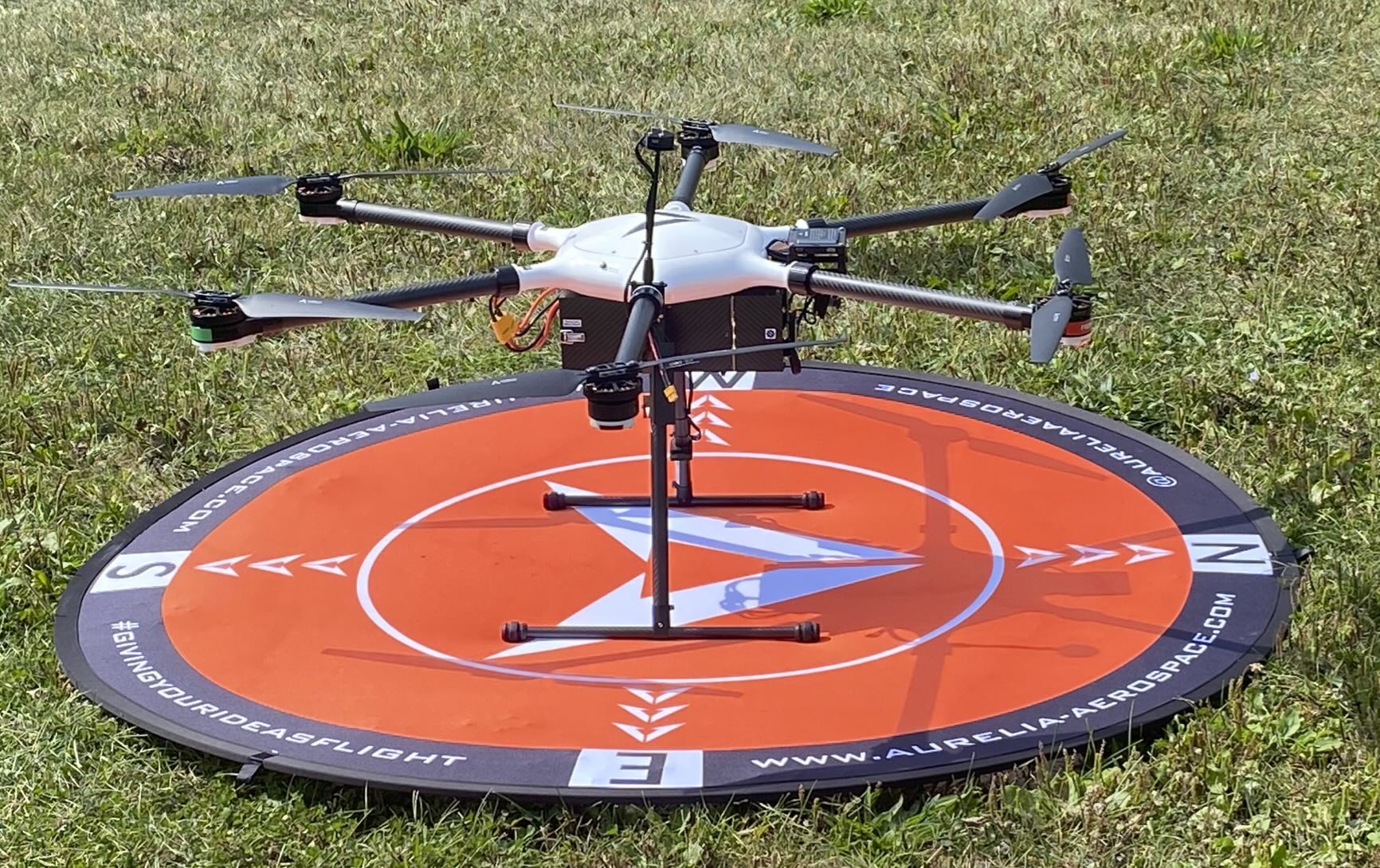}}
\caption{\small In this paper, we explore the potential of heavy-lift drones for Astronomy with an interdisciplinary team of undergraduate and graduate students using two existing University of Michigan drones, including  the Aurelia X6 Pro pictured above.
}
\label{fig:cover} 
\end{wrapfigure}
Unmanned aerial vehicles (UAV) have revolutionized many areas of scientific research, including meteorology, wildlife tracking and monitoring, remote sensing with thermal infrared cameras, and close-up access to dangerous environments such as active volcanoes. Astronomy however has hardly benefited from UAV opportunities due to many factors, such as mass payload limits, short flight times, intense vibrations, and complicated regulatory environment. If these issues can be overcome, UAVs could lift small telescopes, laser or radio beacons, and other optics above ground-level turbulence, permitting low-cost and novel experiments.

Here we focus on commercial off-the-shelf (COTS) multi-rotor drones most often used to carry high-quality cameras for professional photography, cinematography and industrial/agricultural inspections (see Figure~\ref{fig:cover}).  These drones typically incorporate modern differential RTK-GPS navigation that promises accurate (few cm level) 3-dimensional positioning, opening up formation-flying use cases such as interferometry.   There are few  published measurements of the as-built performance for COTS platforms in metrics critical for astronomy, such as vibration power spectra, vibration isolation effectiveness,  precision pointing using OTS gimbals, and real-world XYZ and angular stability under closed-loop operation with internal sensors.   This information is critical to evaluate the feasibility of applications in astronomy, and this paper will provide a first (incomplete) look at important performance metrics.

\begin{wrapfigure}[15]{r}{4.0in}
\centering
\vspace{-.4in}
\mbox{
\includegraphics[width=4.0in]{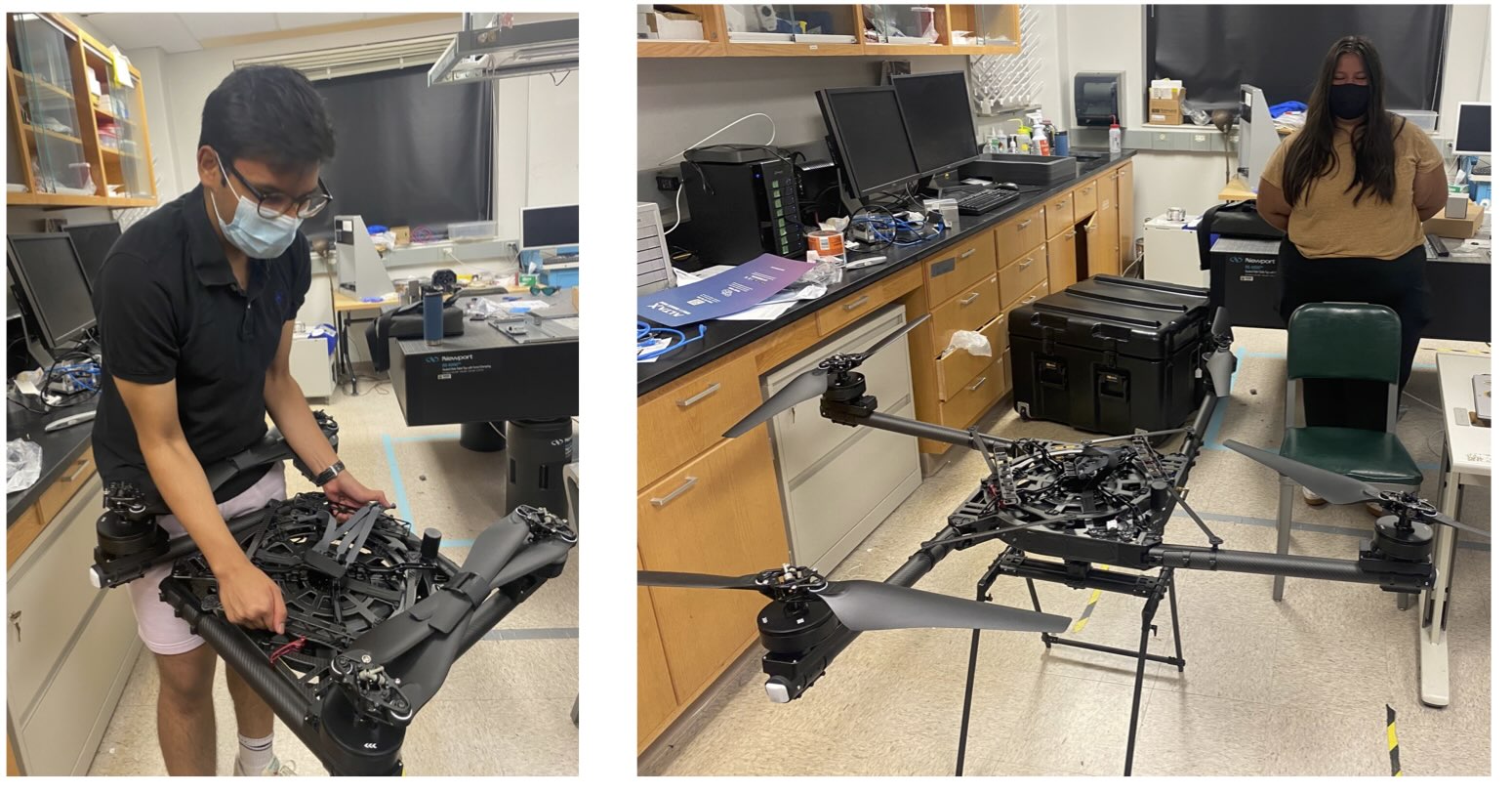}}
\caption{\small The Freefly Alta X is the largest drone we have tested, capable of carrying a payload of 15kg.  While approximately 5~feet across when deployed, the system folds up for compact storage. 
\label{fig:unpack} }
\end{wrapfigure}
 Modern ``heavy-lift" drones can carry up to 15 kg payloads for $\sim$30 minutes with cm-level relative positional knowledge and sub-degree-level angular orientation knowledge.  In addition to characterizing the vibration and stability performance mentioned already, we have begun to measure the effectiveness of vibration isolation strategies using on-board accelerometers while tracking drone motions and rotations with an external monitoring system as a truth test.  For many astronomy applications, diffraction-limited imaging is critical and we have also started the design of a 10cm telescope on a drone to establish the best achievable imaging with arcsecond-level pointing feedback.  

Over the past 3 years, an evolving team of undergraduate and graduate students, drawn from Astronomy, Robotics, and Aerospace Engineering at the University Michigan, has been systematically studying and addressing the main technical hurdles of using commercial off-the-shelf multi-rotor drones for astronomical applications. This report offers a snapshot of our drone research, including a number of characterization tests not published before elsewhere.  We show how existing drones could serve as an integral testing platform for proposed CubeSat interferometers, where end-to-end testing over distances $>100$m is not possible in a static lab setting. Lastly,  we will comment on the feasibility for long baseline interferometry using two formation-flying drones outfitted with laser beacon metrology and tracking cameras.

\section{Drones in Astronomy}
\label{applications}

While this paper is focused on interferometry applications for drones, this section contains a brief overview of  other astronomy-related use-cases.  We hope our characterization studies will be of use to a wider set of instrumentalists interested in exploring drone Astronomy.

\subsection{Atmospheric probes}

Real-time measurements of atmospheric properties can be of great utility for astronomical observations.  For instance, laser guidestar adaptive optics (LGAO) rely on light from a laser beacon to relay information on the turbulence to allow for a realtime correction using a deformable mirror \cite{Wiz2006}.  UAVs are also widely used for meteorological, such as wind measurements using drone-mounted LIDAR \cite{Vasiljevic2020}.  

Instead of a laser guide star, Townes\cite{Townes2002} proposed {\em in situ} measurements of density fluctuations at the ground-layer using backscattered LIDAR radiation to sense path delay as a novel form of ground layer ``adaptive optics'' (GLAO) for interferometry.  This concept can be improved upon by flying a drone with a retroreflector for measuring path length using a ground-launch metrology laser. While the vibrations and motions of the drone itself would need suppressed, either physically or in postprocessing using accelerometer telemetry, this would offer a new technique for realtime ground-layer monitoring. 

\subsection{Imaging above GL turbulence}
Another impactful science case might be to fly small telescopes above ground-layer (GL) turbulence at poor observing sites.  It has been studied that ground-layer turbulence is as significant as high-altitude turbulence for creating bad ``seeing'' image quality \cite{Racine2005}, especially at poor sites.  
Recently published detailed measurements\cite{Butterley2020}  at the world-class Paranal site in Chile suggests significant improvements to ``natural'' seeing if telescopes could be lifted above the $\sim$100m surface layer.  This is well within the technical (and legal) altitude limits of heavy-lift drones and should be pursued.
Another application could be to fly small telescopes on drones to avoid scintillation noise. Ground layer turbulence is the dominant source of photometry fluctuations for small (few cm) telescopes \cite{Dravins1998}.  

One might be concerned that the propellers themselves might induce local turbulence that could ruin image quality for a drone-based telescope.  Fortunately, there have been detailed measurements of this effect.  A Schlieren test was carried out \cite{RodriguezGarcia2019} using a 15.2" DJI propellers placed above, below and within an optical beam.  They found no measurable effect within their precision (rms wavefront disturbance $<\frac{\lambda}{13}$). This reminds us that bad seeing comes from mainly from temperature variations not bulk air motions themselves.  

\subsection{Drones for Calibration of Ground-based facilities}
Drones can be used to calibrate the performance of a variety of  ground-based astronomical facilities. For instance, a drone-based calibration system is being developed to characterize the polarization response for CMB facilities, such as the Simons Observatory SAT telescopes (HOVER-CAL\cite{dunner2020}).  Drone systems have   also been proposed for the Belgian Institute for Space Aeronomy (BISA) radio facility \cite{lamy2014} and  for a mountaintop radio neutrino detector located on difficult site in Greenland \cite{curtisginsberg2021}.  These applications could increase if comprehensive drone flight characteristics were better known.

\subsection{Drones for Quantum Communication}
\label{quantum}
Perhaps some of the most exciting applications for drones are not strictly astronomical, though with potential astronomy applications one day \cite{khabiboulline2019,brown2023}. Drones can be deployed as mobile platforms within quantum networks to enable distributed quantum computing, sensors, entangled resources and  ultra-secure communication. Drones
have been recently flown with laser links between them \cite{liu2019,quantum2021}, experiments sharing many of the same challenges drone astronomy face.

The drone characterization data we publish and the technology we develop will directly benefit these on-going calibration activities and also enable novel synergies with the exploding field of Quantum Communications.

\section{System-level testing of CubeSat interferometry using drones}
\subsection{Why drones?}
\begin{wrapfigure}[17]{R}{4.4in}
\centering
\vspace{-.25in}
\includegraphics[width=4.4in]{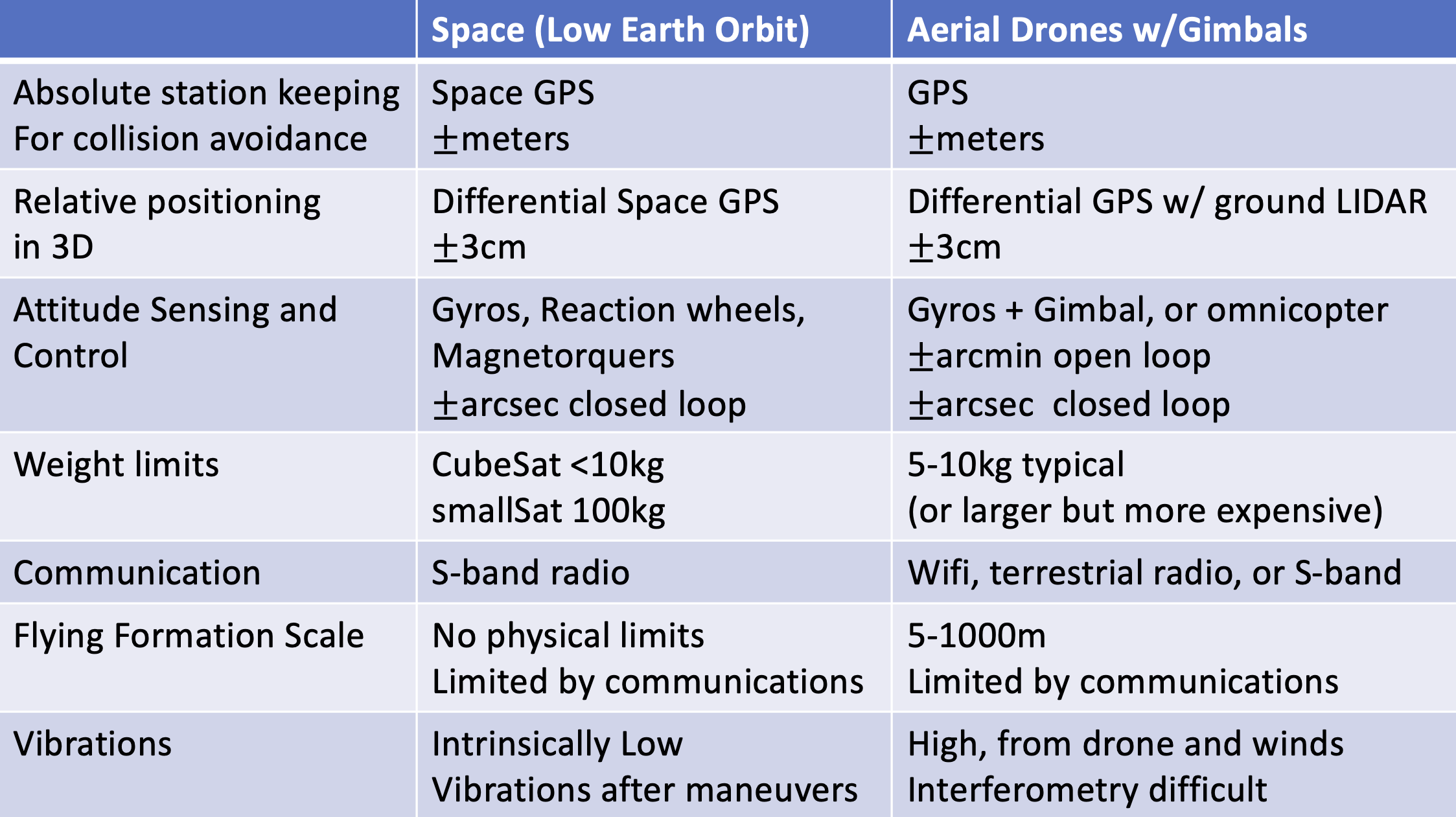}
\vspace{-.15in}
\caption{ We summarize key specifications required for Integration \& Testing (I\&T) of formation flying CubeSats.  
\label{fig_drone_specs}}
\vspace{-.05in}
\end{wrapfigure}
Any proposed space interferometer would fly with much less technical risk following extensive ground-based integration and testing (I\&T).  While many enabling technologies have been demonstrated on a component, or sub-system, level, formation flying itself is difficult  to realistically simulate in the lab.  A science-capable space interferometer will likely need to change baselines from 10s to 100s of meters and inhabit 3 spatial dimensions and control 3 attitude angles.  Furthermore, there will be multiple interacting levels of control loops (tip-tilt, piston, xyz station keeping) that must work autonomously with robust spacecraft safety protocols.  

We have looked to see how other groups have tried to tackle the I\&T challenges for a space interferometer.  In the 2000s, NASA-JPL developed the ``Formation Control Testbed'' (see \url{http://dst.jpl.nasa.gov/test_beds/} for more detail), a formation 
flying lab with 6-dof robots moving on a plane with freely-rotating ``spacecrafts.''  They developed advanced simulation software validated with this real-world lab, albeit with the severe space limitations.

More recently, M. Ireland (ANU) has started building the  Pyxis testbed (\url{http://www.mso.anu.edu.au/pyxis/}) that uses portable and moving robots to simulate 6-dof motions of CubeSat telescopes reflecting starlight to an immobile central station for true star tracking and interferometric beam combination. The use of ground-based robots has limitations in terms of power and baselines, and the motions are constrained in xyz and pointing.

We develop here an I\&T solution for a CubeSat interferometer using the disruptive technology of aerial drones.  While drones may seem at first more complex than conventional robots, they  a) have natural 6-dof motions  without expensive additional actuation, b) can be deployed almost anywhere in wide ($>$100m) formations, c) are relatively inexpensive due to large commercial markets, d) allow robust system-level testing even within high-vibration environment.

\subsection{Flight technologies addressed (and not addressed) by the Drone Platform}
\begin{wrapfigure}[16]{R}{4.4in}
\vspace{-.25in}
\includegraphics[width=4.4in]{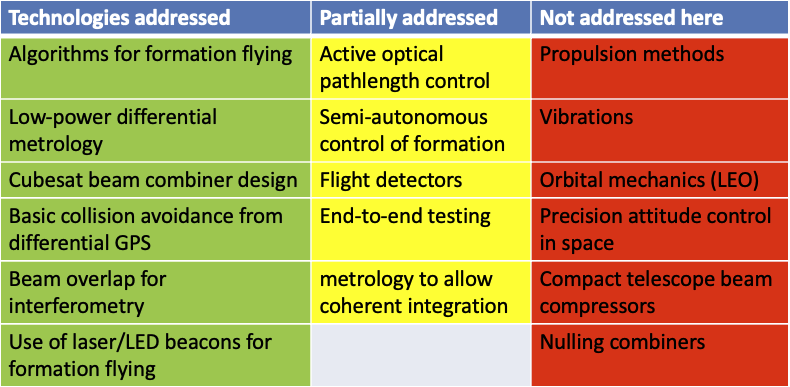}
\caption{The Technology Readiness Level (TRL) of formation flying algorithms and subsystem components can be addressed by the Drone Platform,  either significantly (green), partially (yellow), or not at all (red). \label{fig_tech}}
\end{wrapfigure}

 From the point of the view of the spacecraft control computers, the drone will respond to xyz commands using rotors much as a CubeSat would respond with cold gas or ion thrusters. Similarly, the rotors $+$ gimbals of a drone will allow closed-loop feedback to a star-tracker or beam-overlap camera much as a CubeSat would utilize reaction wheels or torque rods.  Figure~\ref{fig_drone_specs} show the ways that drones can mimic the space environment for CubeSats, with approximate specifications. Station keeping, attitude control and weight limits are all comparable between space CubeSats and commercial drones, though we expect vibration environment to be far worse on a drone.  Even here, there is a commercial market for stabilized high-magnification drone photography which means that both passive and active technologies exist to mitigate the worst of the vibrations.

As already discussed, we believe this approach will retire risk and optimize chances of future flight success. That said, not all important CubeSat interferometer technical challenges can be addressed and we have summarized these in Figure~\ref{fig_tech}. Specific tasks that our project can address: xyz station keeping,  beam transfer by controlling precise pointing of telescope payloads, precision beam steering, compact beam combiner design, multi-agent coordination, low-power differential metrology,  use of LED beacons for relative navigation, and  establishing algorithms and control loops for precision formation flying of 3 spacecraft. We will not address propulsion, precise CubeSat pointing in space, orbital mechanics, design of the telescope beam compressor for larger collecting area, nulling combiner designs, vibrations/fast  control of optical path length, or full optical truss using 2D metrology to allow long coherent fringe integrations.

\section{Overall Concept for Drone Interferometer}

\begin{figure}[!t]
\centering
\includegraphics[width=6.5in]{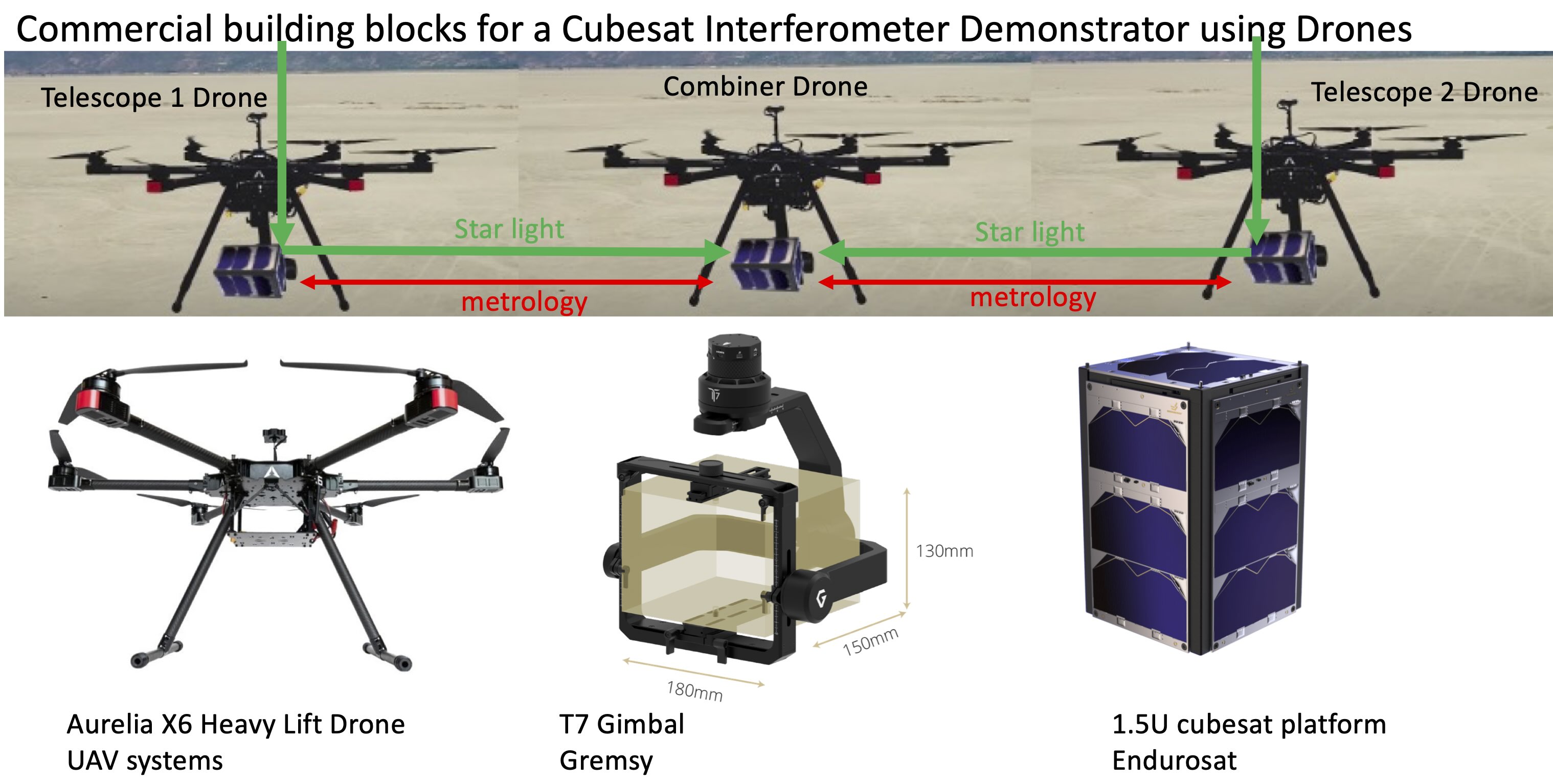}
\caption{Here is a visualization of the formation flying concept using gimballed hexacopters.
We show here commercially available aerial drones (Aurelia X6 Pro), a gimbal system from Gremsy, and a 1.5U CubeSat platform from Endurosat. Image credits to vendor websites: \url{http://uavsystemsinternational.com}, \url{gremsy.com}, \url{www.endurosat.com}
\label{fig_drones}}
  \vspace{-.1in}
\end{figure}

We describe a novel  testing platform using commercial aerial drones for testing formation-flying CubeSats.  As discussed, drones provide flexible 6-dof control of a CubeSat payload over potentially 100s of meters of separation.  We hope to eventually carry-out end-to-end testing of a 3-drone interferometer system, including tracking an artificial star, reflecting the light from two telescopes to a central combiner, and injecting into an interferometer combiner.  While we will monitor the differential pathlengths with laser metrology and maintain beam overlaps in closed loop, our initial goal is not stabilized interferometry due to expected high vibrations of the drones, although we discuss potential to suppress vibrations in \S\ref{longbaselines}.   A mockup of the system can be seen in Figure~\ref{fig_drones} .

\subsection{Hardware Details: Telescope, Metrology and Beam Combination}
\label{hardware}
In order to understand the building blocks of the interferometer architecture, we will follow the starlight through the proposed system and discuss each subsystem in turn.

\begin{figure}[!t]
\centering
\includegraphics[width=6.5in]{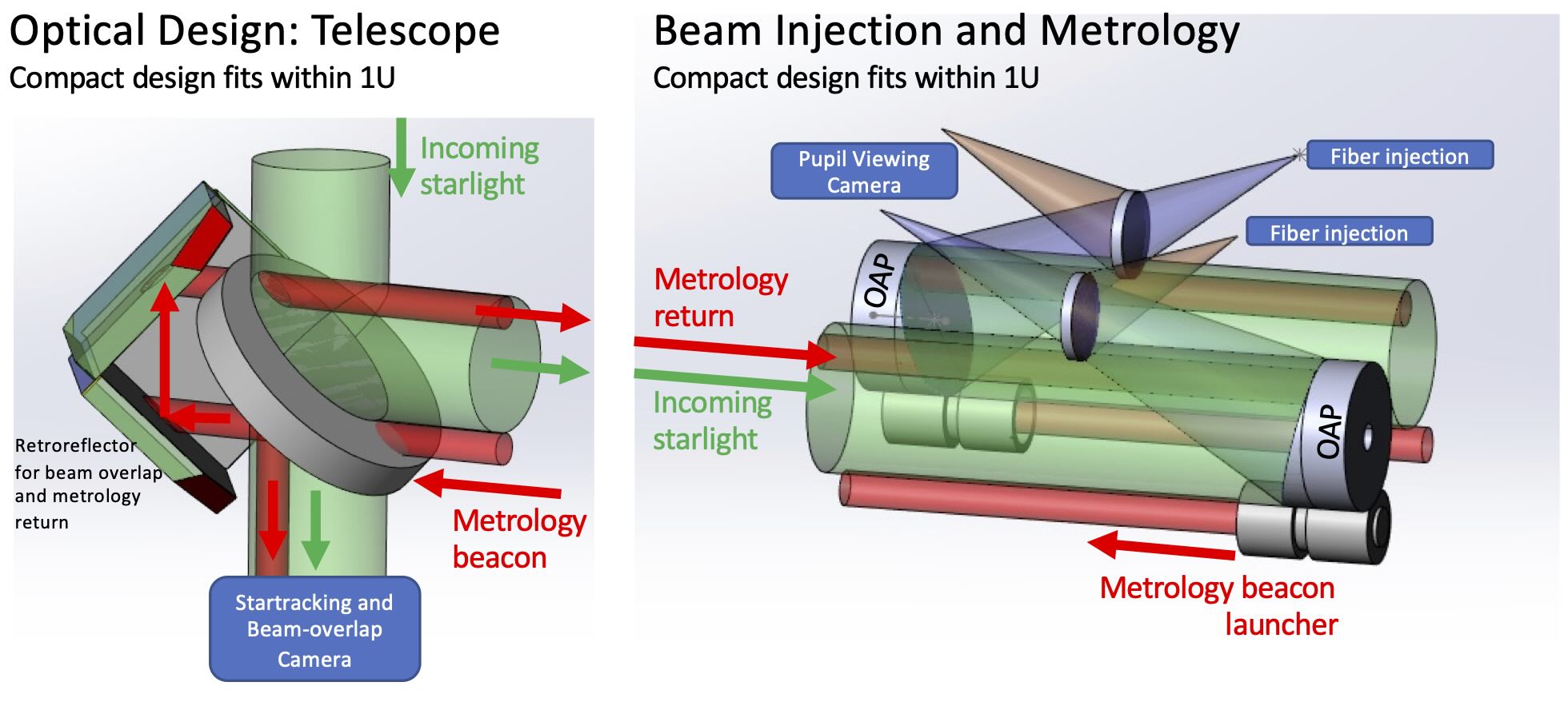}
\caption{(left) This shows the preliminary optical design for the telescope and combiner drone payloads. See text for a guided tour through the optical train.  
\label{fig_optical_design}}
\vspace{-.1in}
\end{figure}

{\bf Telescope Unit:}
Incident light will be from either a real star or artificial light source. A small siderostat on each drone will reflect a 25mm beam of light towards the combiner drone.  Figure~\ref{fig_optical_design} (left panel) shows a CAD drawing of the incident light (green rays from top) reflecting off a 90:10 beamsplitter at a 45$\arcdeg$angle of incidence.  For greater collecting area, this optical design could be replaced with a 100mm aperture beam compressor to increase collecting area while still forming a 25mm beam for the combiner.  In this prototyping stage, we will use the drone gyroscope telemetry for coarse angle pointing and use a startracking camera behind the beamsplitter for pointing feedback.  

One challenge for formation flying interferometry is the diffraction-limited  overlapping required for the telescopes beams entering the interferometric combiner.  Here, we solve this problem by launching laser metrology beacons from the combiners that are co-aligned with the fiber injection off-axis parabolas (OAP).  Thus, by using a retroreflector, we can measure precise beam overlap at each drone and implement fast tip-tilt correction here using the gimbal feedback directly or using PZT actuators.   We are aware of the strict optical tolerances this puts on the co-alignment of the OAP and metrology launchers but this level of precision is common in optical interferometry instrumentation on the ground which has to contend with strong atmospheric turbulence.

{\bf Combiner unit}:
We now follow the star light and the metrology return beam from the telescope toward the combiner (right panel, Figure~\ref{fig_optical_design}).  The retroreflector has translated the original metrology beam and now both starlight and metrology come to a common focus for injection into a suitable single-mode fiber. As drawn, both sets of 25mm f/4 OAPs  will fit within as small a volume as a 1U CubeSat when placed side-by-side.  In this design, we can introduce a beamsplitter to monitor the pupil position.  Our notional design places fibers on same side of spacecraft to minimize fiber length and bends on its way to the beam combiner.

The type of beam combiner for a future CubeSat mission will depend on the priority science focus, though we anticipate single-mode fibers could be used due to their stability, robust alignment methods, and potential for a range of combiners including optimal nullers using integrated optics \cite{hsiao2010,martinod2021}.  For initial work, we plan to use a simple image-plane combiner similar to ones used in the MIRC, MIRC-X, and MYSTIC combiners\cite{mirc2004,anugu2020,setterholm2023}.  This inexpensive system has no moving parts and will have sufficient spectral resolution to isolate the starlight from the metrology fringes.  We will use a high speed CMOS detector in hopes of following atmospheric turbulence seen by the metrology system and to potentially detect white light fringes as well (in ``lucky imaging''\cite{law2006} mode).

{\bf Metrology unit:} The final piece of our system is the metrology system.  Previous interferometer concepts (TPFI, SIM) and previous laser metrology smallSat missions (i.e., GRACE-FO) used conventional long coherence-length lasers to monitor path length fluctuations between the combiner and telescope arms.   Such systems are not compatible with the mass, power, and volume constraints of CubeSats and we adopt a new compact Fiber Bragg Grating (FBG) Laser (see \S\ref{metrology_prototype}) with 3m coherence length.  By sending the same metrology beam out to each telescopes and then back, we can interfere this return light and operate near zero optical path difference.  Other schemes are also possible\cite{lagadec2020}.   
Internal metrology is probably optional for a space interferometer as a natural guide star can be used for fringe tracking. However, here the metrology system will also provide valuable characterization of the vibration and formation stability.  We discuss our metrology desing in more detail in \S\ref{metrology_prototype}.

\subsection{Multi-agent coordination and control}
The whole point of using drones for I\&T is that getting all the interferometer systems working smoothly is complex and requires end-to-end testing under dynamic and variable conditions.  The challenge of this task should not be under-estimated,  especially in light of the fact that the formation must be able to fly semi-autonomously without humans in the control loop for extended periods of time.  This is of course true in space where radio communication is intermittent but also on the ground with drones due to safety issues, the multiple interacting layers of control (xyz, gimbal, beam overlap, pupil control, etc.), and the high speed of the control bandwidth.  We believe the lessons learned building drone flight software will translate into better CubeSat flight software by confronting real-world experience, not just controlled simulations and tests within confines of a lab.  The area of semi-autonomous multi-agent coordination is ubiquitous across the diverse field of robotics, beyond aerial drones or space, and should be addressed with modern software infrastructure and tools.

\section{Characterizing the performance of two commercial heavy-lift drones}

\subsection{Description of Individual Drones}

While there are many types of UAVs that could be considered for astronomical purposes, here we limit our discussion to multi-rotor drones. Multi-rotor drones typically have 4--8 propellers and come in a variety of sizes.    The largest commercially-available multi-rotor drones have mass limits of 5--15~kg (beyond its own weight), sufficient for a modest scientific payload.  Most drones of this class can be tightly integrated with a gimbal to smoothly direct a camera at any angle, typically for photography/cinematography and site inspections.

\subsubsection{Multi-rotor Drones}

In early 2022, we surveyed the marketplace for drones with the following characteristics:
\begin{enumerate}
\itemsep =0em
    \item Minimum payload mass 5kg
    \item RTK-GPS capable with differential XYZ precision $\pm$3cm
    \item Option of {\em top-mounted} gimbal to allow starlight acquisition, with arc-minute open loop pointing precision
    \item Flight time $>$30 minutes
    \item Cost $<$\$25K
    \item Open source software controller and compatibility with standard mission planning software
\end{enumerate}
\begin{wrapfigure}[14]{L}{3.6in}
\centering
\vspace{-.2in}
\includegraphics[width=3.5in]{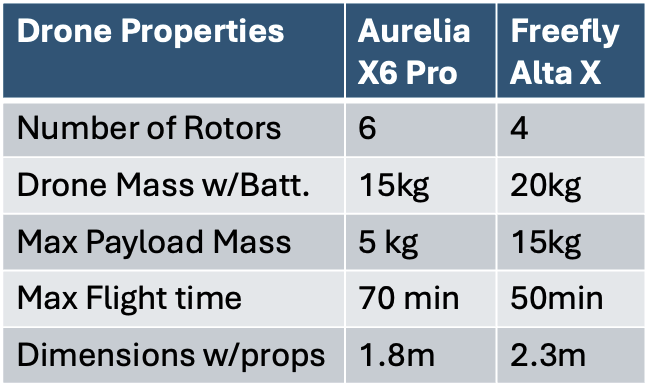}
\caption{{\small The basic characteristics of the drones under test. 
\label{fig:dronetable}}}
\end{wrapfigure}

Based on thorough web searches of specialty drone websites and discussion boards, we followed up with technical support for detailed information on the Freefly Alta X series, UAV Systems Aurelia X6 Pro, and DJI Drones. 
The first two of these met our specification while DJI drones were unable to offer top-mounted gimbals or sufficient mass limits.  We summarize relevant technical capabilities of our chosen drones in Figure~\ref{fig:dronetable} with pictures in Figures~\ref{fig:cover} \& \ref{fig:unpack}.

All of the multi-rotor systems we investigated were based on fixed-orientation rotors in a plane.  This means that to move laterally, or to stabilize against winds, the entire system has to tilt to produce sideward forces.  This is undesirable for formation-flying interferometry application we discuss in \S\ref{longbaselines} as the drone payload translates and rotates while maintaining fixed sky position.  We also considered ``omnicopters,'' a class of drone with multiple rotors either arranged in 3D and/or with steerable propellers. These additional degrees of freedom allow the drone payload to be oriented in 3 spatial and 3 angular degrees of freedom without an additional gimbal.  While we preferred this option for the formation-flying use case, there were no COTS omnicopters available with sufficient lift capacity, and early discussions with Department of Defense and Swiss researchers building omnicopters did not lead to a viable collaboration.

In this paper, we will present preliminary characterization data collected over the past 2 years.

\subsubsection{Drone Accessories}

Drones come with many potential accessories.  In order to have stabilized filming of ground targets while flying, most commercial drones offer gimbals.  A top-mounted gimbal with a camera is shown in Figure~\ref{fig_gimbaltraining} and discussed later in this paper.  Readibly-available gimbals, for example ones by Gremsy, offer roughly arcminute pointing on a (passive) vibration-isolating stage.  The best software (e.g, from Freefly) combines the gyroscope data from the drone body plus from the gimbal to allow a smooth closed loop guidance for the most demanding filming, capabilities needed for Hollywood movies. 

While available drone gimbals are stable enough for filming, they are not well-suited for some astronomy applications. For instance, a telescope mounted on most gimbals, with representative camera cage dimensions 15cm x18cm x13cm, would not have room for an instrument at a traditional Nasmyth or Cassegrain focus.  Further, for optical links needed for interferometry or quantum key distribution,   the light beams often must be redirected at 90~degrees towards a second drone.  In general, the redirected beam will run into the axis drive of one of the gimbal motors, for conventional designs.  Lastly, filming requires just arcminute precision while the diffraction-limit of a small 10-20cm aperture is at the arcsecond level.  This demands a 2nd stage tip-tilt mirror or a more precise gimbal for astronomical use.  We discussed our specifications with industry experts at multiple companies and determined a custom half-yoke alt-az or alt-alt design will be needed astronomy purposes (see more discussion in \S\ref{altaz}).

Other common accessories available on most drones are ground-directed LIDAR for better altitude precision for landing and collision avoidance, RTK-GPS base stations for cm-level differential XYZ positioning, cameras for remote viewing while flying, flight simulators for safety training of operators, extra batteries, and software tools for mission planning.  

Lastly, we investigated the use of tethers for these drones. Tethers can be used (and are sometimes required) as purely a passive safety measure to avoid a runaway drone endangering humans or property. Another interesting application for astronomy is using a tether as a power and data conduit.  With a powered tether, a drone can run indefinitely through an electrical connection and can send data at high rates using fiber. Further, one could even send laser signals or other information to the drone for specific experiments, such as a local oscillator needed for phase-coherent interferometry or certain quantum communication experiments.  The tether will add mass to a flying drone that increases with altitude but will also provide stable power for maximum sustained thrust for hours or days.  We do not further consider tethers here but note the importance for some future applications.

\subsubsection{Regulatory Environment}

Flying heavy-lift drones pose a serious safety risk to humans and property.  The Federal Aviation Administration (FAA) regulates the operation of drones and the University of Michigan has additional layers of rules.  The regulatory environment is dynamic and astronomers working with drones will need to stay alert to any changes, such as the flight ban of DJI drones recently threatened by Congress.

For completeness and reader interest, we include some basic guidelines that impact the use for astronomy:

\begin{itemize}
\itemsep=0em
\item Each drone must have an individual operator who has passed an FAA online training course and must not leave the operator's line-of-sight
\item Drone must stay below 400 feet from ground level and 
stay 8km away from local airports.
\item Drones may operate at night but with certain lighting requirements that may contribute light contamination for astronomical payloads. 
\item The requirement to carry a registered Broadcast Remote ID has recently been expanded to include all pilots flying out of FAA-recognized identification areas.
\item Some FAA requirements can be waived for higher education/research purposes (upon application and approval).
\item In addition, many universities (including the University of Michigan) forbid all outdoor drone flights on university property and indoor flights can be subject to further rules, such as use of a restrictive safety tether and a documented and reviewed safety plan. 
\end{itemize}

\begin{figure}[!ht]
\centering
\includegraphics[width=6.5in]{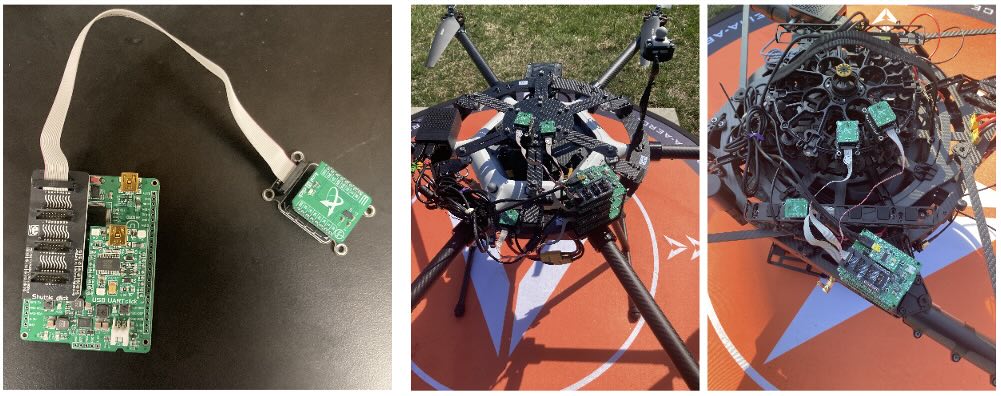}
\caption{(left) Picture of the accelerometer and microcontroller used in our vibration experiments.  (middle) Here we see the accelerometer locations for the Aurelia drone and  (right)  for the Alta X drone. 
\label{fig_vibration1}}
\vspace{-.1in}
\end{figure}
\subsection{Vibration Measurements}
\label{vibration}
Anyone who has flown a drone knows that fast-turning rotors introduce strong vibrations on the body of the drone at frequencies of many kilohertz.  The level of vibration is not constant with time and can vary substantially based how balanced the propellers are -- no specific balancing was done for this study and we used the drones as they came from manufacturer.  Our group found few published papers\cite{verbeke2016,verma2020active,Ishola2022} with quantitative measurements of the vibration spectrum and the effectiveness of common isolation stages.  While not exhaustive, we present our initial studies of vibrations on heavy-lift drones.  

We purchased 3 MEMS accelerometers to measure and compare vibrations on different parts of the drone. Our accelerometers are ST IIS3DWB, which are 3-axis sensors with 75$\mu g/\sqrt{Hz}$ sensitivity and 6 khz bandwidth. We utilized many components from Mikroelectronika, our accelerometers are mounted on Accel 14 Click boards (MIKROE-4185) and attatched to mikro bus shuttles (MIKROE-3785) to connect our accelerometers to PIC-32 microcontrollers (MIKROE-1717). This system allows us to place accelerometers anywhere around the drones for our vibration analysis (see Figure~\ref{fig_vibration1}).

\begin{figure}[!b]
\centering
\includegraphics[width=6.5in]{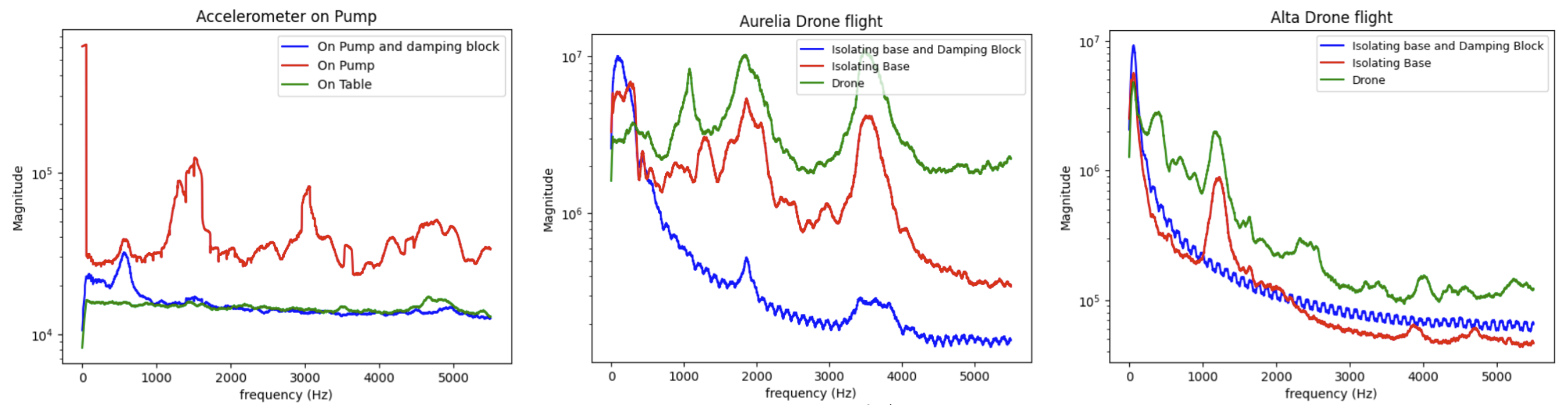}
\caption{Here we see magnitude spectra of the 3 accelerometers for three experiments described in the text: (left) reference measurement in the lab near a vacuum pump, (middle)  Aurelia X6 Pro drone data during a flight, (right) Freefly Alta X drone data during a flight.
\label{fig_vibration2}}
\vspace{-.1in}
\end{figure}

We used the``mikroC Pro for PIC-32'' compiler for programming our microcontroller. Our program polls the accelerometer card and buffers the data in chunks of 100 data points before sending the values out as bytes through a USB port connected to a Raspberry Pi. Careful testing was done to assure maximum bandwidth without telemetry dropouts.  The Raspberry Pi converts the binary data to strings and saves the results in csv file to analyze after flight.   The Raspberry Pi can be mounted to the drone body and powered by auxiliary battery cables offering voltage conversion.

Since it is difficult to calibrate the magnitude of our vibration spectrum in an absolute sense, we first took a reference measurement.  We used a lab vacuum pump as a source of vibration and placed accelerometers on and around the pump. We know from experience that we can carry out interferometer experiments on the surface of this optical table while this pump is running. 
We placed an accelerometer on the optical table next to the pump, one on the pump mounting plate, and another on a Sorbothane vibration damping pad on the pump mounting plate. Figure~\ref{fig_vibration2} (left panel) shows the average 'magnitude spectra' of the 3 accelerometers.  From the magnitude spectrum of the pump vibrations, we see that Sorbothane isolates high-frequency vibrations well above 1 khz, but Sorbothane is not as effective at lower frequencies as expected. We also see that a vibration magnitude of $\sim$2$\times10^4$ (on this arbitrary scale) can be considered as quiet, with vibrational amplitudes $\Delta X << 1\mu$m.   This means that interferometry experiments could be done on drones if we can isolate vibrations to this level.

Figure~\ref{fig_vibration2} also contains the vibration spectra for flights on the both drones.  
We mounted the three accelerometers on the Aurelia and Alta X drones as shown in Figure~\ref{fig_vibration1}.  We placed one on the main body of the drone, another on the drone's vibration-isolating platform (buffered by rubber), and the last one on a Sorbothane damping pad we affixed to the vibration-isolation platform.   We also attached the microcontrollers and a Raspberry Pi computer to the drone body in order to control the accelerometers and store the vibration data during flight. We could communicate with the Raspberry Pi using a wireless WIFI connection in realtime while the drone is flying.  

From inspection of the accelerometer power spectra show in Figure~\ref{fig_vibration2}, we can draw some conclusions.  1) Both drone bodies have similar high vibrations at low frequencies $<$ 1 khz, with the Alta X vibration power fulling off substantially up to 4Khz while the Aurelia drone maintains high frequency power  with strong peaks.  The springy-rubber isolation stages and the Sorbothane pad both have some good suppression effect reducing amplitudes by a factor of 2-5, though only at frequencies above 500Hz.  We see the Sorbothane pad further reduced high frequency vibrations on the noisier Aurelia drone by another factor of 5 but this material only had  a marginal affect for the quieter Alta X. 

For the Aurelia we are 100$\times$ above quiet levels at high frequencies and $\sim$20$\times$ higher for the Alta X. The larger physical size and slower rotor speeds for the Alta X might explain the better performance, as well as the ``Alta ActiveBlade" design which claims 5 times vibration reduction compared to other propeller designs, which is close to what we observed in fact.  

We have only investigated the simplest of vibration isolation strategies so far, ie, rubber springs and Sorbothane pads.  We are optimistic that more advanced isolation strategies could attain the required stability for interferometry and suspect this is a solved problem in the military context.  See more discussion in \S\ref{longbaselines}.

\begin{figure}[!b]
\centering
\includegraphics[width=6.85in]{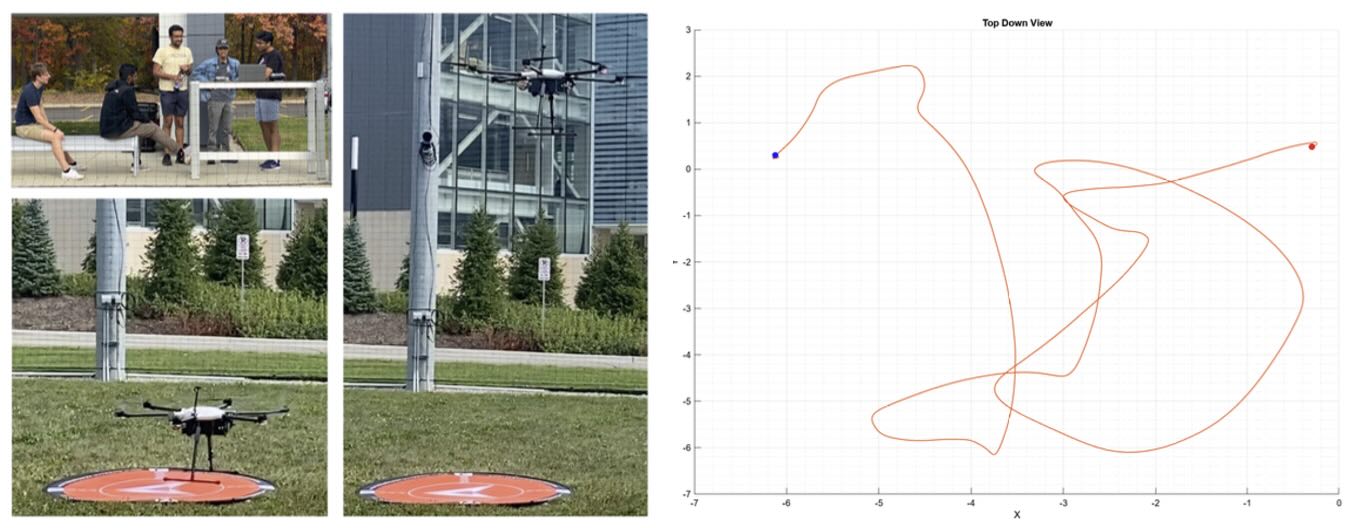}
\caption{(left) Our flight stability tests were carried out at the M-air facility at the University of Michigan.   (right) Here we see an example track of an early drone flight, from the top view, using the M-air built-in motion capture technology.
\label{fig_stability2}}
\vspace{-.1in}
\end{figure}

\subsection{Drone Flight Stability}

The ability for the combiner and telescope drones to fly together in a precise formation is  necessary to ensure that star and beacon light can be transmitted between the two drones. The drones must be able to record data about its own position to inform the other of its position, from which the receiving drone must adjust its position to maintain alignment. 
Each drone has specific flight modes that need performance testing to determine how to attain best flight stability.  For the telescope drone, the modes tested for stability were position hold, altitude hold, and loiter.  Our team used the University of Michigan M-air facility to characterize the flight stability of our drones.

\begin{figure}[!b]
\centering
\includegraphics[width=6.85in]{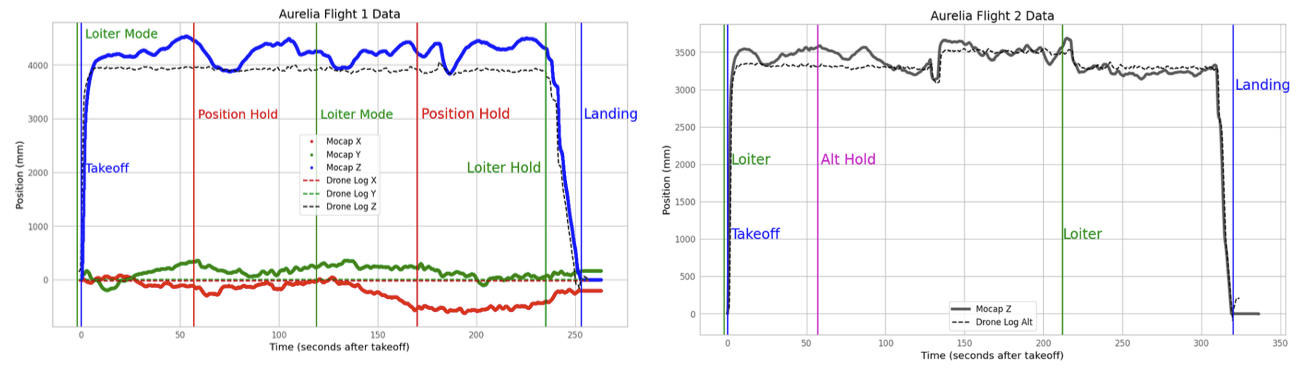}
\caption{Here we see the internal position telemetry from the Aurelia X6 Pro drone (dashed lines) compared to the actual position (solid lines) from a Motion Capture system  with $\pm$1 mm errors). (left panel) Flight one show testing of Position Hold and Loiter modes in all three (XYZ) dimensions.  (right panel) Flight two shows just the altitude dimension using the Altitude Hold mode.  For these tests, stand-alone GPS was used without a base station.
\label{fig_stability3}}
\vspace{-.1in}
\end{figure}

M-air (\url{https://robotics.umich.edu/about/mair/}) is a 10,000 square foot, four-story, fully netted outdoor enclosure on the Engineering Campus designed to allow safe experimentation of prototype hardware with no risk to humans or property.  In addition, M-air is outfitted with a powerful array of cameras allowing full 3-dimensional realtime tracking of a drone's flight.  This motion capture (MOCAP) system observes an array of reflective spheres on each drone to allow monitoring of position, attitude, and orientation.

The MOCAP system provides an external ``truth test'' that allows us to validate the onboard positional data derived from GPS, gyroscopes, and accelerometers. Through careful analysis comparing the drone log data and motion capture data, we can determine the true behavior of the drones in each of its available modes and use that knowledge to make a decision on what mode can provide the most stability to enable precise formation flying.

We present early testing of the Aurelia X6 Pro in  Figure~\ref{fig_stability3}.  
The first flight began at 7:34 PM on November 13, 2022, with a takeoff lasting seven seconds to reach an altitude of  about four meters, then hovering. This test was an uninterrupted hover with testing of multiple flight modes.  The figure below depicts the behavior of the drone in the X, Y, and Z axes during the flight and we label each flight mode as we switched between ``Loiter'' and ``Position Hold'' modes.
We can immediately identify a few issues with the drone and its understanding of its own location. To begin with the most glaring error in the drone log, we can see that the altitude of the drone varies up to half a meter from what it believes that its position is at certain times in the flight. This can also be seen in the recorded video footage of the drone flight when no control is given but the altitude changes nonetheless. As GPS is generally less accurate in the z-axis, we might not be surprised by this result (we later added a ground-looking LIDAR sensor which drastically improved the altitude flight stability).  Looking at the X and Y data, we can see that the drone MOCAP position deviates only slightly at first from the log data but then beings to drift away from its original position in the X direction, which correlates with an eastern drift. Overall, the drone believes that it is almost perfectly still in both modes but in fact there was a considerable amount of drift, roughly 20-40cm.

When looking at the different modes, we see in flight 1 that we take off in loiter mode and then make changes between loiter and position hold throughout the flight. By taking the standard deviation of each axes over the mode intervals, we find that loiter mode provides a more stable flight for altitude, but position hold provides more stability for the X and Y dimensions.  In flight 2 (right panel of Figure~\ref{fig_stability3}), we tested ``Altitude'' hold mode which did indeed show the best stability in the altitude dimension compared to Loiter or Position Hold. 

Our first flight of the Freefly Alta X flight was in 2022 and ended in a small crash (no damage), which later was determined to be pilot error.  These large heavy drones do not respond quickly to commands, and the time to stabilize following take-off is longer than for smaller drones. Our first successful flight was on April 9, 2023 and we report results here from this day.   
In order to ensure a safe takeoff, both flights started in Manual mode, which allows the pilot to command roll, pitch, and yaw angles while offering rudimentary stabilization. This drone uses a different drone controller and we tested performance of ``Manual'' vs ``Position Hold'' modes.   We found position mode was much more stable than Manual mode, although we can not present MOCAP data for these flights -- MOCAP requires operation at night but our testing here was during the day.  Note that we also flew the Aurelia drone this day and visually found the Alta X to be much more stable, showing fewer erratic jumps and less drifting.  The larger size and inertia of the Alta X should make the platform more robust to wind impulses and its more advanced (dual frequency) GPS might show smaller drifts (see \S\ref{gps}).

Overall, this analysis allows us to understand which flight mode is ideal for each drone. For the Aurelia, this is generally Loiter mode, and for the Alta X, it is position mode. These modes allow the pilot to command an exact position, rather than a roll rate. The drones then use all of their sensors to hold this position, regardless of wind or drift. However, we have found that sensor information is the limitation to high fidelity and that the observed level of drift ($>$20cm) would not be precise enough to allow light to be reflected from one drone to another.  In order to overcome this limitation, we next investigate the accuracy and precision of differential RTK-GPS using a fixed base stations to correct for common-mode GPS errors.

\subsection{GPS telemetry accuracy}
\label{gps}
In the last section, we confirmed using the MOCAP data that stand-alone GPS has errors exceeding 20cm. While expected, this is not sufficient for our formation flying requirements since we need to hold formation to with a few centimeters, the size of the collecting optics in our experiment.  Fortunately, both drones are equipped with RTK-GPS capabilities, which allows the use of a 2nd GPS base station to be used as a local reference.  By taking the realtime difference between the drone GPS and base station GPS, the relative position of the drone should be now accurate to a few centimeters according to drone manufacturer claims.

Due to technical problems with the motion capture system at that time and the logistical difficulties of nighttime experiments, the following tests were carried on the University of Michigan central campus with the propellers off.  We wanted to test the accuracy and precision of RTK-GPS by simply moving the drone on the ground in a series of exactly 2 meter steps.  By carefully moving the drone in a known pattern, we can then compare the internal drone position telemetry with the truth steps to gauge accuracy. We refer to this test procedures as the ``Step Test.''

First we tested the Freefly Alta X system.  The Alta X produces .ulg files, which must be reduced using a third-party converter. To do this, we found a python script called ulog2csv (\url{https://github.com/PX4/pyulog}), which produces a machine-readable comma separated (csv) spreadsheet of telemetry data. Upon running the script, the file would separate into dozens of different files which individually contained data for actuator control, battery status, wind estimates, and more. We were focused on position, for which we found 3 relevant files: local position, global position, and GPS position. The global position and GPS position files are recorded in degrees latitude and longitude, while the local position was recorded in meters. Since our target precision is on the order of centimeters, we converted the global position and GPS position to meters using a conversion factor of 111139 m/degree longitude. As our experiments had the drone moving North and South, we did not have to consider longitude.

\begin{figure}[!t]
\centering
\includegraphics[width=6.85in]{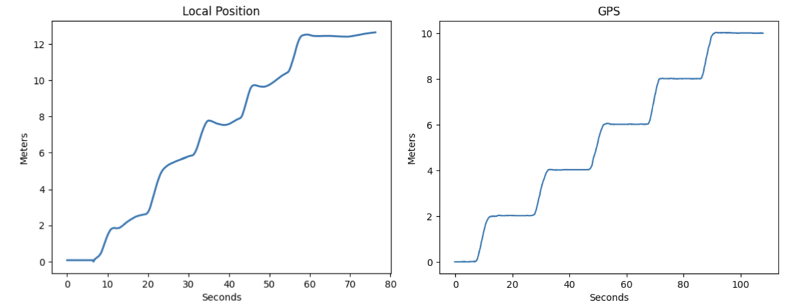}
\caption{Freefly Alta X GPS ``Step Test'' results. Stand-alone GPS (left) vs differential RTK-GPS using a base station  (right).  We see excellent results using RTK-GPS, with erorrs $\Delta X<2cm$, sufficient for maintaining an interferometer formation. 
\label{fig_gps1}}
\vspace{-.1in}
\end{figure}

The left panel of Figure~\ref{fig_gps1} shows the internal drone telemetry for the Freefly Alta X system during a representative ``Step Test,'' without the assistance of the external GPS ground station. Purely qualitatively, this plot shows only a vague indication that the drone was moving in a “step” pattern. There is no clear stopping and starting points at each interval and it instead seems to simply slow down, or have a brief change in direction when it actually stopped completely for 10 seconds. 
Quantitatively, the system measures that it traveled far beyond the 10 meters it actually moved, going all the way past 12 meters. While it is hard to pick the exact point where the drone stopped moving based on this data, the ranges it traveled between each step also vary significantly, ranging between $\sim$1.5 meters and $\sim$3 meters, whereas the actual distance it traveled was 2 meters between each step. 

The right panel of Figure~\ref{fig_gps1} shows the internal drone RTK-GPS position when the ground reference GPT base station was used.   We see a clear “step” pattern where the drone moves to a new location, stays there for 10 seconds, and then moves on to the next location.  The original GPS data was collected in degrees of latitude/longitude, and we converted to meters using the conversion factor described above. The rms variation when not moving was only $\pm$2cm and the distance between each ``step'' closely matched the expected 2 meters.    We performed two more trials of the step test with the Alta X (with the GPS ground station), and found similar results each time.

\begin{figure}[!b]
\centering
\includegraphics[width=6.85in]{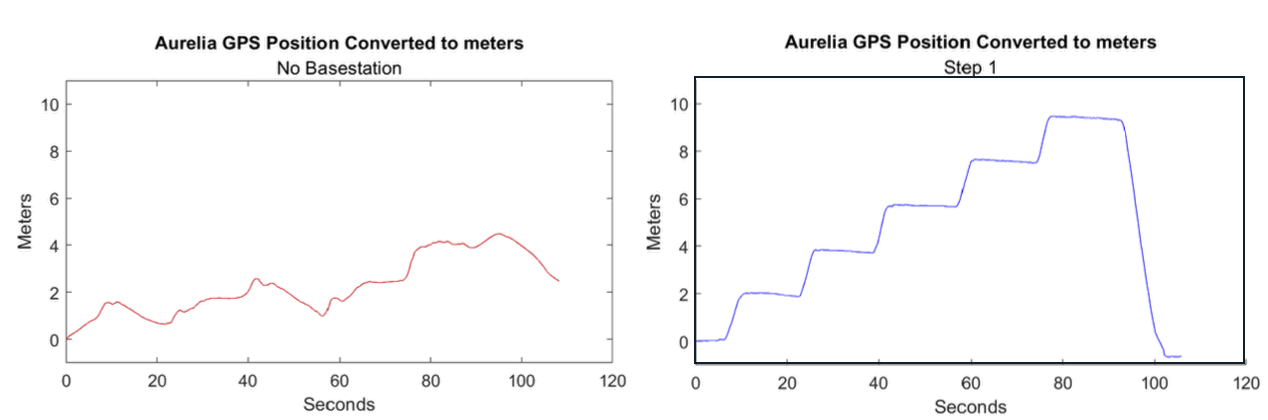}
\caption{(left) Aurelia X6 Pro ``Step Test'' results. Stand-alone GPS (left) vs differential RTK-GPS using a base station  (right).  We see noticeably poorer results than for the Alta X drone. While the RTK-GPS base station provides radically improved position precision, we see unexplained linear drifts that significantly degrade accuracy.
\label{fig_gps2}}
\vspace{-.1in}
\end{figure}

We now present results from the Aurelia X6 Pro, which uses a different control system and so the analysis proceeded slightly differently.  This drone produces .bin files, which also need converted to a more useful format.  Thus, we needed to find some conversion tool that worked specifically for the files from Aurelia drones produced by the Ardupilot software onboard. We were able to find a third party converter called Ardupilog (\url{https://github.com/Georacer/ardupilog}), which converts the .bin files to one large Matlab structs. Much like the Alta X data, this struct contained dozens of data fields. We found the most relevant fields to be GPS and IMU acceleration. We were particularly interested in the IMU acceleration as it may have provided some insight into some drifts we saw. However, as we will outline shortly, the IMU acceleration did not have a clear affect on the positions errors.

Figure~\ref{fig_gps2} (left panel) shows  the position data collected for the Aurelia X6 pro during the step test without the assistance of the external GPS ground station. The data shows no indication that the drone was moving in any kind of “step” pattern. The recorded data shows the drone moving no farther than 5 meters, which is significantly different from the actual distance of 10 meters. This is clearly unsatisfactory and inadequate for precision formation flying, thought fortunately the results using RTK-GPS base station were far better.

Figure~\ref{fig_gps2} (right panel) shows the position data collected for the Aurelia X6 Pro during the step test with the assistance of the external GPS ground station. The series of steps is very evident now, with an rms error of $\pm$7.1cm when not moving.  However, in this step trial, we can see noticeable positional drifts at each step that were not present for the Alta X data. At the first step, for example, the drone starts at a position of 1.94 meters, and right before it moves to the next location it records a position of 1.87 meters (0.07 m difference). Every step saw a similar level of drift, with the maximum occurring at the fourth step, where the drone started at a position of 7.66 m and ended at a position of 7.51 m (0.15 m difference). Taking the average of the change in position due to drift at each step, we found that the drone recorded an average change of 0.11 m per step. Likely because of this drift, the overall distances traveled had significant error as well. The final step (before the negative drift) recorded a position of 9.44 m, 56 centimeters short of the actual position of 10 meters. 

Interestingly, we found the overall ``drifting'' rate was not constant but changed with time.  Later repeat ``Step Tests'' showed sometimes minimal drifting but this was not repeatable even after waiting for the electronics to settle after a drone reboot.  We looked into the IMU telemetry to see if there was a correlation but did not find one. 
 On the plus side, the system seems to provide precise values on short time scales, but the large drifting will need to be eliminated before we can use the Aurelia X6 Pro for precision formation flying. Currently we are considering two explanations: a) the Aurelia used a lower cost, single frequency GPS system which could explain differences, or b) the reported GPS positions we are using are not correct for some reason, due to misunderstanding of the telemetry log parameters.  We will work with manufacturer to ensure we are interpreting the telemetry correctly and achieve the best possible performance.

{\em In conclusion, we have verified sufficient positional accuracy and stability to carry-out a formation flying experiment using the Freefly Alta X drone.}  We believe the Aurelia X6 Pro can also be within specifications, if we can understand the source of the steady ``apparent'' positional drifts seen in our telemetry.  This may require using different GPS modules or finding a software explanation in the way the accelerometer and GPS data are fused in the control system.

\section{Sub-system testing}
\begin{wrapfigure}[23]{R}{4.0in}
\centering
\vspace{-.15in}
\includegraphics[width=4.0in]{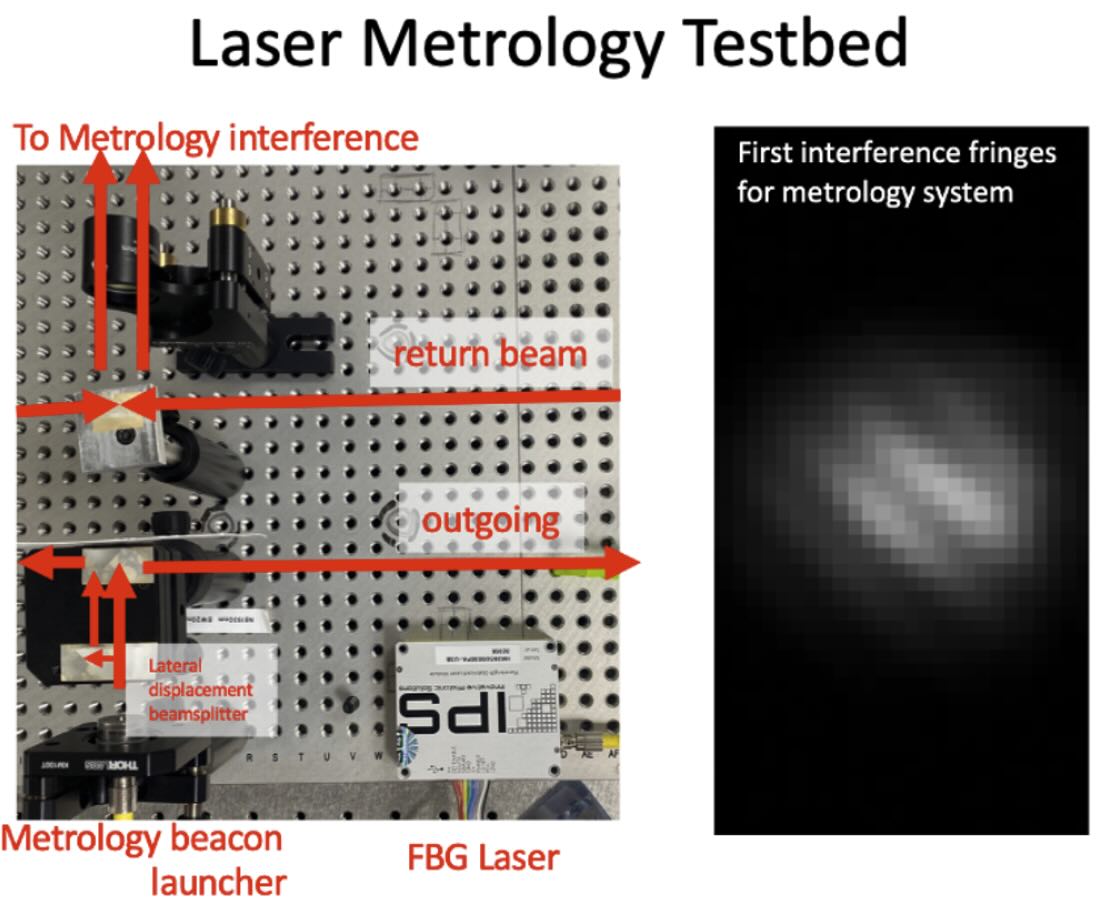}
\caption{{\small In order to calibrate absolutely our MEMS accelerometers and for future potential interferometry applications,  we have designed a compact metrology payload (left) and first lab fringes with the system are shown (right).  
\label{fig:metrology}}}
\end{wrapfigure}
In addition to the characterization tests shown above, UM students worked in the lab to prototype drone interferometer subsystems. Here we report progress on the Metrology Testbed, LED beacon testing, Gimbal tracking, and the design of a new drone telescope mount.

\subsection{Metrology testbed}
\label{metrology_prototype}
While space GPS will allow us to know the relative positions of our spacecraft within a few centimeters, we need to monitor the separation at the precision of a wavelength of light for space interferometry.  To do so, the central combiner spacecraft will send out light from a laser diode that will reflect off the two “telescope” spacecrafts and return for interference, allowing a differential path length to be monitored.  Missions such as the GRACE-FO (\url{https://gracefo.jpl.nasa.gov/}) have already carried out high precision metrology between two spacecrafts but we will be doing so in a much smaller package that can be hosted on a cubesat.  In \S\ref{hardware} and Figure~\ref{fig_optical_design}, we already discussed a scheme for launching the metrology beacon and for recombination.

In the context of drones, realtime measurements of metrology could help to compensate for vibrations and positional drifts, although fringes from stars or artificial light sources could also provide this information if bright enough.   Here we built a prototype of low-power, inexpensive, and  compact metrology system that could fly on a drone or fit within a cubesat (see Figure~\ref{fig:metrology}). 
Note this system is also a useful tool in the lab to calibrate our accelerometers on an absolute scale.

The laser we use here is a small (0.1kg) Fiber Bragg Grating (FBG) laser diode integrated with a thermistor and thermo-electric cooler from IPS (I0638SU0030PA-USB), with an instantaneous  bandwdith $<$100MHz and coherence length of $>$3m.  In our testbed, a fiber launcher is used to create a small metrology beam which is split into two parallel beams using a lateral displacement beamsplitter. The beams hit a rooftop prism and are sent in opposite directions toward the two telescope drones.  
 The laser beams in each arm return after reflection off retroreflectors mounted on the telescope drones.  The two light beams are brought back together with another rooftop prism and an imaging lens onto a high speed ($>$1kHz) CMOS detector.  Figure~\ref{fig:metrology} shows  an image of our testbed along with first interference fringes.

 We do not have flight data to report at this time.

\begin{figure}[!b]
\centering
\includegraphics[width=6.5in]{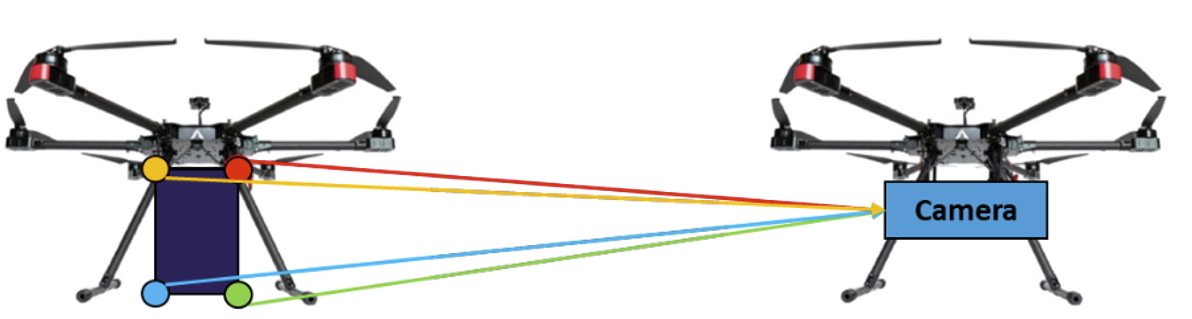}
\caption{We will use LED beacons and a camera to estimate relative positions of our drones/spacecraft to within 1 millimeter, augmenting the cm-level positions available from differential RTK-GPS.
\label{fig_led_concept}}
\vspace{-.1in}
\end{figure}
\subsection{LED beacon testing}

The ability for the combiner and telescope drones to maintain a precise distance and orientation is of utmost importance to successful beam transfer between the drones. While coarse positioning and orientation will be handled through the GPS system, the visual beacon system offers an additional layer of position and orientation information to the two-drone system. Using a set of LEDs on one drone and a camera on the other, we can deduce the relative distance and orientation of one drone with respect to another. This information will then be fed to our control loop to adjust the relative positions and orientations in real-time.  Camera images are commonly use in space rendezvous and docking procedures and LED beacons were previously used in the PRISMA\cite{bodin2012} formation flying mission.  

A schematic of our concept is shown in Figure~\ref{fig_led_concept}. While RTK-GPS capabilities provide positioning accuracy within 2~cm (see \S\ref{gps}), our goal for the beacon system will be to provide an accuracy on the order of millimeters for position in the X, Y, and Z  axes.  We imagine a constellation of different colored LEDs on each cubesat that will allow the orientation, scale factor, and position to be measured with a fixed focal length camera.

\begin{wrapfigure}[23]{R}{3.8in}
\centering
\vspace{-.0in}
\includegraphics[width=3.8in]{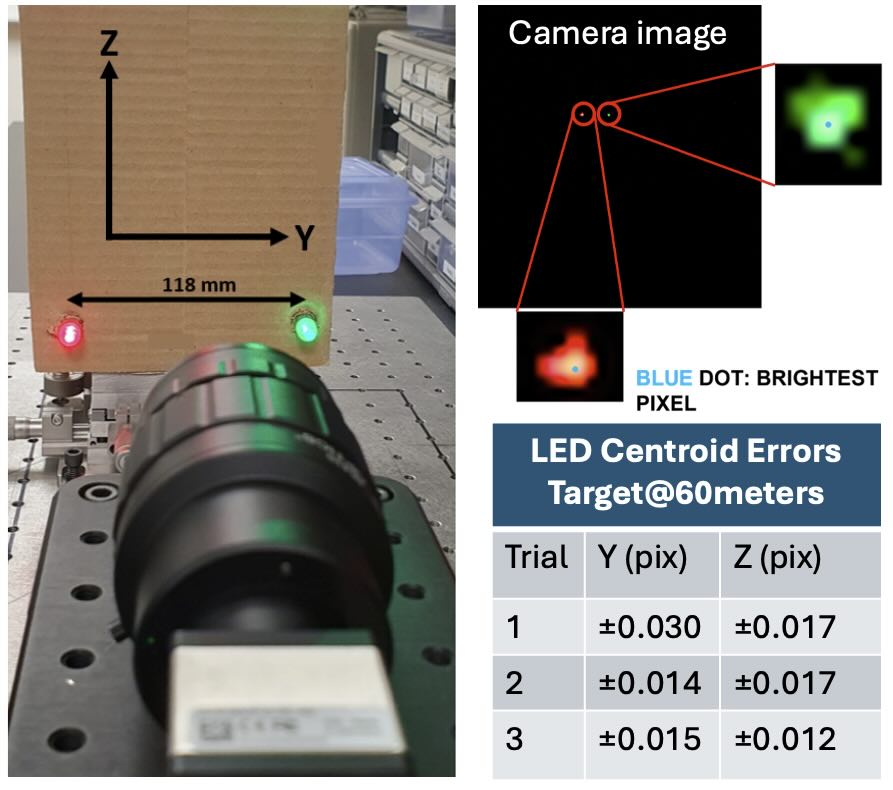}
\caption{{\small Left panel shows the lab setup of our LED beacon test. Top right panel shows image with a zoom up on the red and green LED images.  We also include a table of our centroid errors from 3 trials, demonstrating sub-pixel precision.
\label{fig:fig_led1}}}
\end{wrapfigure}
As an example of our procedure, consider a 100 meter baseline separation between spacecraft.  Using a camera with a f=150mm focusing lens,  LEDs separated by 10cm would be imaged as spots separated by about 3.4' (arcminutes), which is 150$\mu$m at the camera's focal plane. For CMOS pixels of 4.8$\mu$m, this is a separation of $\sim$31pixels.  If the camera has a 10mm aperture then the diffraction-limited spot full-width-at-half-maximum (FWHM) will be about 12'' (arcseconds), or about 2 pixels.   In order to have 1 millimeter resolution, we would need to measure each spot at 2'' precision, or about 1/3 pixel.    With a high  signal-to-noise  ratio (SNR) image, we can easily measure image centroids with an error smaller than a pixel -- formally, the centroiding error $\Delta\Theta \sim \frac{\Theta_{\rm FWHM}}{{\rm SNR}}$.

To demonstrate this concept with realistic hardware, we imaged two colored LEDs separated by 118mm using a CMOS camera and a f$=$100mm lens in the lab. By locating our experiment in a very long hallway, we were able to separate the camera and the LED beacons by 60~meters.  The setup and results of this experiment are shown in Figure~\ref{fig:fig_led1}, finding typical centroid errors of $<$1/20 pixel, far exceeding our requirement. We moved the LED using a micrometer and confirmed that we could track the motion accurately.  Furthermore, we confirmed large photon count rates even when using LEDs blocked down to small pinholes and reading out the CMOS sensor at high frame rates ($>$10Hz), important for freezing drift motions in a more realistic dynamic environment.   

Note that in order to interpret image positions as a relative spacecraft position, then we will also need to know separately the attitude position of each spacecraft at the $\simle$arcsec level in 3 axes of roll, yaw and tilt.  While challenging, this can be done using two precision star trackers oriented at 90 degrees to each other.

\subsection{Gimbal Tracking}
\label{gimbal}
\begin{figure}[!t]
\centering
\includegraphics[width=6.5in]{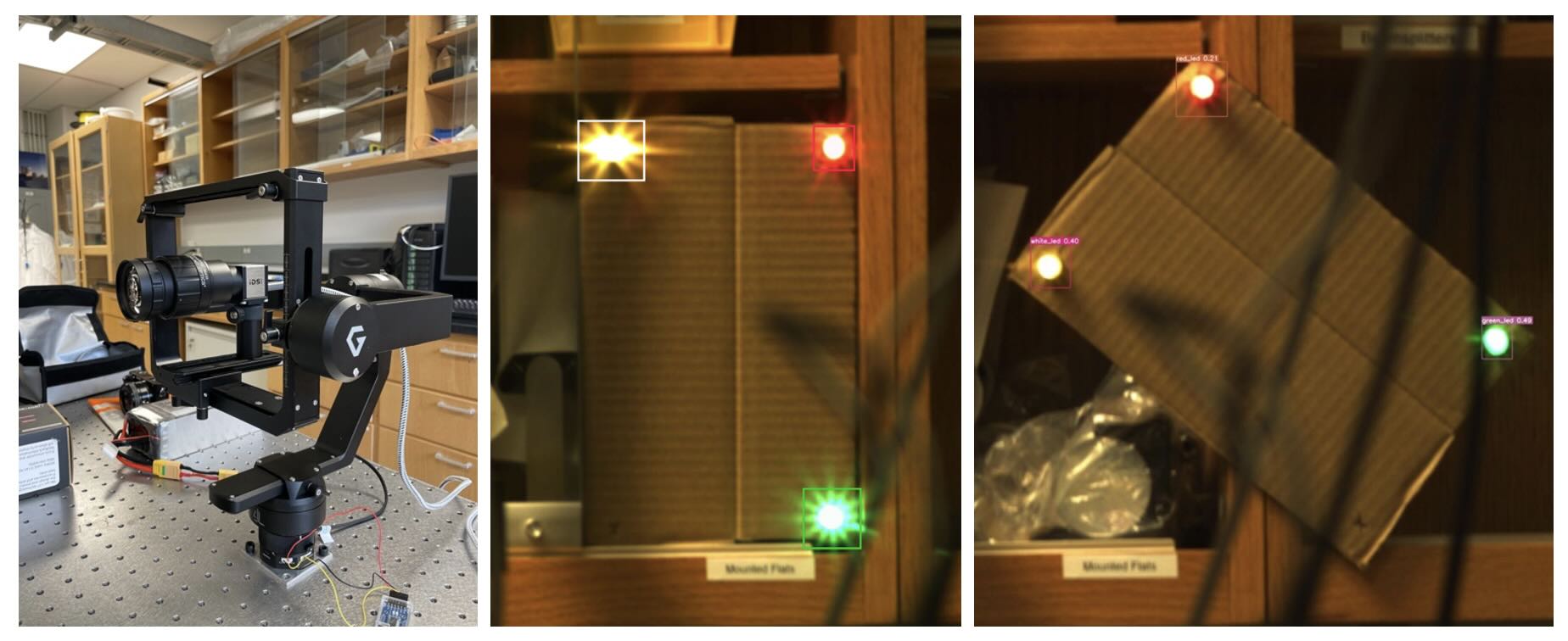}
\caption{(left) Here we see the Gremsy T7 gimbal mounted on a HDMI hyper quick release adaptor in the lab for testing.  A camera has been mounted and balanced in the gimbal for tracking experiments.  (middle) A machine learning algorithm was trained to recognize the LED patterns using training images with rectangles annotations around the required LEDs. (right) Here we show the result of positive identifications on test images by the machine learning algorithm as marked by the overplotted rectangles with correct color labels and corresponding confidence scores.
\label{fig_gimbaltraining}}
\vspace{-.1in}
\end{figure}

Our drone interferometer testing program requires a gimbal to track the LED system above during flights.
Figure~\ref{fig_gimbaltraining} (left)  shows an image of a IDS CMOS camera on the Gremsy T7 gimbal system in the lab. The Gremsy T7 gimbal can be controlled using a serial interface operating at $\sim$10Hz with an encoder precision of $\sim$1' (arcminute).  As will be discussed in the next section, this level of attitude control is not adequate for drones separated by $>$100m baselines but is adequate for initial experiments to reflect light from one drone to another while flying in formation over short baselines of ~$\sim$5~meters,  as is possible in our netted M-air facility. We can mount this gimbal to fly on either drone using the Gremsy ``HDMI Hyper quick release'  interface.  

Controlling the gimbal using a Raspberry Pi coomputer was more difficult than expected.  The Gremsy T7 gimbal is normally controlled through the Ardupilot drone flight software. This is convenient for conventional filming while using a manual joystick pointing to control the camera.  In our case, we want to operate the gimbal in closed-loop with a tracking camera at video frame rates.   Thus, we must bypass the normal communication channels and have a direct connection to the gimbal through the USB port of a Raspberry Pi computer. 

Following wiring suggestions at \url{https://github.com/Gremsy/gSDK}, we used the COM2/COM4 port (connector JST SM06B-GHS-Tb), which uses the UART serial protocol.  A commercial TTL-to-USB adaptor (DSD TECH SH-U09C with FTDI FT232RL) was needed to connect to the Raspberry Pi.  With the new cabling and installation of Gremsy Firmware v780 Beta through a Windows helper application,  
we then downloaded the open Software Development Kit  at \url{https://github.com/Gremsy/gSDK/tree/platform-test} onto our Raspberry Pi Debian system. This SDK, written in C++, contains predefined libraries that are specifically tailored for controlling the gimbal.  We noted many incompatibilities between certain firmware versions and different gSDK branches, but found a good  combination of the platform-test branch with firmware v780 beta.  Gremsy customer support was helpful working through these difficulties. 

We successfully compiled this code on the Raspberry Pi and it was able to run a test program  'mavlink\_control.cpp', which demonstrated example gimbal control and status commands. We could then modify this source code as the basis of our custom application.   We plan to incorporate a socket server into this C-based control system to allow other applications -- such as the python image analysis algorithm described next -- to interface with the gimbal for live tracking.  

With the gimbal control working, we move on to configuring the IDS imaging camera. We installed the available linux driver IDS UI Rev 2.1,   which allows our cameras to be readout using python on the same computer that controls the gimbal.  Here we also ran into some difficulties finding compatible drivers versions,  as the Raspberry Pi Debian kernel is not the same as for desktop PCs. 
We successfully installed PyuEye which is a Python package allowing us to configure and readout the camera.

For a space mission where the camera will be viewing one cubesat, we can likely utilize a simple image processing algorithm based on gaussian fitting of the LEDs spots.    However, the drone use-case is more complicated since the camera will always see a more complicated and ``busy'' scene with confounding objects.  For instance, the area around the M-air flight facility has many building lights and even a stop light in the field-of-view. Thus, we have begun to look into more advanced image recognition algorithms to apply for our tracking system.

We implemented a machine learning algorithm trained to spot the LED beacons in a complex image scene while not falsely triggering on other lights or objects.
While this is a ``work in progress,'' we show initial results using the YOLOv7 (You Look Only Once) algorithm which utilises neural networks. The YOLOv7 algorithm excels with simple objects and offers the advantage of being computationally efficient to permit real-time operation.  The first step required us to train the neural network using images of the beacons. We captured images at five different orientations of the LED beacons, which here contained 3 different LEDs.  We then divided the training dataset into positives and negatives. The positive dataset included images in which all three LED lights were clearly visible. In contrast, the negative image set consisted of images that either displayed only a partial view of the beacon configuration or did not contain the beacon at all. The algorithm employs the negative images without changes, but the positive images require rectangle annotations to pinpoint the locations of the lights, as shown in Figure~\ref{fig_gimbaltraining} (middle). For annotation, we used CVAT, an online computer vision annotation tool. Figure~\ref{fig_gimbaltraining} (right) shows the example of successful beacon identification using the YOLOv7 algorithm.

While automated tracking of LED beacons in closed-loop has not been demonstrated yet, we have commissioned the building blocks, including control of the gimbal control through C++ and the python image analysis using a lab CMOS camera. These routines all run concurrently within the same Raspberry Pi computer that can be installed on a drone, as we have already done for the vibration test (see \S\ref{vibration}).  

\subsection{Drone telescope upgrade study}
\label{altaz}

In this subsection, we summarize our efforts to design a custom telescope mount that could work for a drone interferometer.    The important specifications that make this a unique problem:
\begin{itemize}
    \item We need tip-tilt control at the arcsecond level with a bandwidth of $\sim$100Hz. This is beyond the capabilities of conventional commercial drone gimbals which typically have arc-minute precision at $\sim$10Hz. 
    \item Unlike a normal focusing telescope, our system is a ``beam compressor" that reduces a 10-20cm beam to approximately 2cm.  Further, the output beam must be reflected 90$\arcdeg$ from the incoming direction towards the combiner drone.   This requires a half-yoke design to avoid obstruction and also specific combinations of axes to be motorized.
    \item To be lifted by a drone, we need to minimize the mass of the mount and use lightweight telescope 
    \item The mount should be integrated within a vibration isolation stage on the drone.
\end{itemize}

One of the most serious deficiencies of COTS drones for astronomy is the limited arcminute pointing stability derived from the available gimbals.  We considered custom solutions from Newport and HDAir, but price estimates were prohibitive. Instead, we began to design our own simple alt-az (or alt-alt) telescope mount based on lightweight mechanics and motors integrated with arcsecond-level precision rotary encoders to allow diffraction-limited pointing.  Following market research and many industry inquiries, we have identified direct-drive motors by MTL Japan to be sufficiently small and with embedded 21-bit rotary encoders to meet our specifications and cost requirements.  

\begin{figure}[!hb]
\centering
\includegraphics[width=6.5in]{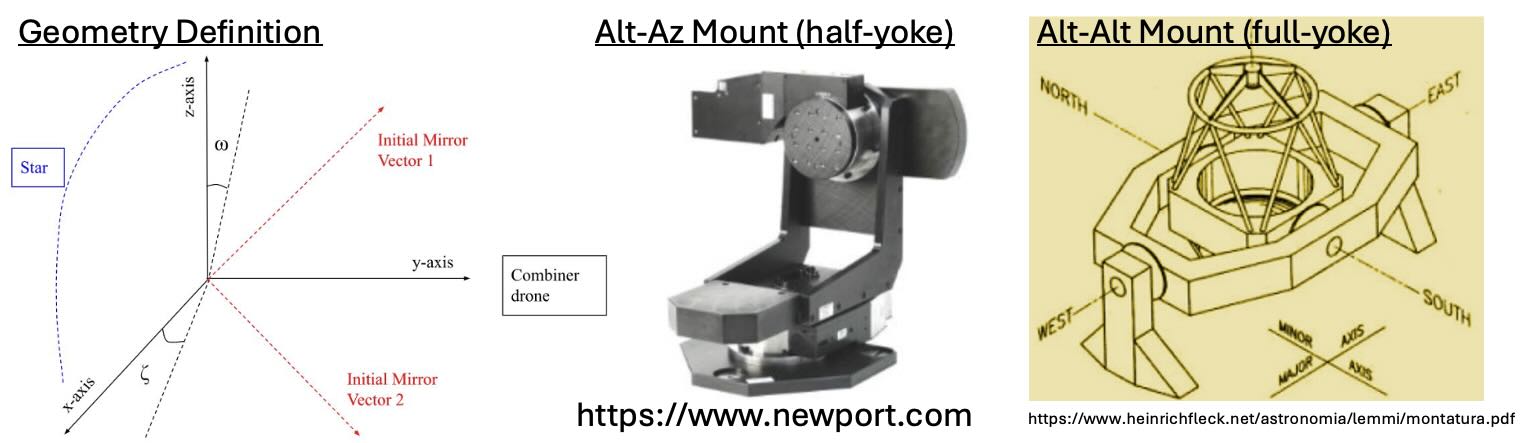}
\caption{We studied multiple telescope mounts that could be used for our application.  Using the geometry defined here (left), we conclude a half-yoke alt-az mount (middle panel) or an half-yoke alt-alt (right panel) would be suitable.  See text for further description.
\label{fig_dronetelescope1}}
\vspace{-.1in}
\end{figure}

The precision motors will be incorporated into a conventional half-yoke, alt-az mount as depicted in Figure~\ref{fig_dronetelescope1}.  This open design will allow telescopes of differing diameters and configuration to be adapted.   We aim to initially outfit this mount with 10cm f/3.4 telescope based after a CubeSat design by M. Ireland (ANU, private communication).  The telescope mass will $<$1~kg, well within our payload limits and suggesting that 20cm versions could be lofted eventually. 

Figure~\ref{fig_dronetelescope1} defines a coordinate system that use here: y axis pointing from telescope drone to combiner, z-axis is up, and x-axis pointing forward horizontally from telescope drone.  
From the mount designs presented, the best choices for this application would be an half-yoke alt-az mount with a y-axis rotation nested in a z-rotation.  Alternatively,  an alt-alt mount with a y rotation nested in an x rotation would also work.  We have not made either of these designs yet, but include illustrations in Figure~\ref{fig_dronetelescope1} (middle and right panels).  

We hope to proceed to a CAD design in the next phase of this project.  

\section{Could we use drones for long-baseline interferometry?}
\label{longbaselines}

One  area where drone astronomy could actually surpass current facilities is for very long-baseline interferometry. 
At first, our team did not consider this possibility as feasible, having extensive real-world experience fighting vibrations on multiple ground-based interferometers (IOTA, Keck-I, CHARA).  However with further analysis, drone-based interferometry might be a realistic possibility after all.  Consider the following points:
\begin{itemize}
\itemsep=0em
\item Lightweight deployable telescopes on drones can be larger than the atmospheric coherence length $r_0$ in the  visible  -- {\em especially in the blue \cite{davis2005} where no interferometers exist now.}  This makes the collecting area competitive with current visible interferometers which all lack visible adaptive optics, e.g., NPOI, CHARA.   
\item Long delay lines and atmospheric dispersion compensators are unnecessary for drone interferometers, since the formation itself can maintain equal light paths within a few centimeters by flying on a baseline roughly perpendicular to the direction of the source.  This is a {\em dramatic} infrastructure savings and would boost the optical transmission by an order of magnitude compared to ground-based facilities by reducing then number of required mirrors reflection from $\sim$25 to $\sim$5. 
\item {\em A key insight is that there is no need to stabilize the drones themselves to within a wavelength of light -- a small delay line locked to a fringe tracker will stabilize slow fringe drifts due to drone wander while stabilizing  fast changes due to atmospheric turbulence. } 
\item Vibrations pose the highest risk for an drone interferometer and it is not immediately obvious if this can be adequately addressed.  While a naked drone indeed suffers intense shaking driven at a few 1000~Hz by imperfectly balanced propellers, vibration-isolation platforms using springs, Sorbothane-like plastics, pressurized rubber bumpers, contact-less voice coil actuators, air bearings, piezoelectrics,  and other technologies could drastically reduce the high-frequencies, and such platforms have been deployed for airborne imaging systems, especially in the military context \cite{verma2020active} but also civilian (e.g., SOFIA\cite{sofia2010}). 
\item {\em A key insight is that vibrations only need to be reduced below the irreducible floor set by atmospheric turbulence, with typical visible light coherence times of a few milliseconds. } We can think of the residual vibrations as a kind of ``extra seeing'' which can be corrected using conventional tip-tilt and fast fringe tracking control loops if not too large. 
\item The science potential for even a single-baseline drone interferometer is clear and compelling.  Drones can easily fly beyond the longest optical interferometer baselines of 330m (CHARA), opening up  measurements of stellar diameters and binary orbits for entirely new classes of objects with angular resolution $<$0.3 milliarcseconds, especially massive, hot stars with high surface brightness that have been too small to resolve before.
\item Drones could also be used for other forms of interferometry: intensity interferometry, heterodyne interferometry using a distributed LO through a tether or laser link, or even at longer radio wavelengths using a dipole array.   Although in these, the small apertures pose a more significant sensitivity limit.
\end{itemize}

While technically risky, a drone interferometer could have a large potential science reward at a relatively small cost.

\section{Conclusions}
Here we have argued for developing multi-rotor drones as a testbed for interferometry, both for integration and testing of CubeSat technology and for attempting stellar measurements. We have presented initial vibration tests find a potential for strong and significant suppression using conventional strategies, with the larger Alta X system showing lowest vibration levels. We also validate the cm-level accuracy for Alta X RTK-GPS system in real-world tests.  Our team hopes to attract additional funding to continue this development work towards more on-sky flight experiments, construction of a robust vibration-isolation stage, a small drone telescope mount, and the purchase of additional drones.

\acknowledgments 
We would like to acknowledge formative discussions with our colleagues Michael Ireland, Jonah Hansen, Kira Barton, Gautam Vasisht,  Saptarshi Bandyopadhyay, Michael Meyer, and Sascha Quanz, who have generously contributed their advice and expertise. We also thank Aurelia Aerospace for extensive customer support.
 
We would like to recognize seed funding from the Michigan Institute for Research in Astrophysics (MIRA) and Michigan Space Institute.    This research utilized the M-air facility, operated by the University of Michigan College of Engineering.

\bibliography{report} 
\bibliographystyle{spiebib} 

\end{document}